\documentclass[doublecol]{epl2} 
\usepackage{graphicx}
\usepackage{dcolumn}
\usepackage{bm}
\usepackage{amssymb}
\usepackage{amsmath}
\begin{document}
\setcounter{page}{1}
\vskip 2cm
\title{The graviton Higgs mechanism}
\shorttitle{\textbf{The graviton Higgs mechanism}}

\author{Ivan Arraut$^{(1,2)}$}

\institute{$^1$ Department of Physics, Faculty of Science, Tokyo University of Science,
1-3, Kagurazaka, Shinjuku-ku, Tokyo 162-8601, Japan \\\\
$^2$ State Key Laboratory of Theoretical Physics, Institute of Theoretical Physics, Chinese Academy of Sciences, Beijing 100190, P R C}

\abstract{The Higgs mechanism at the graviton level formulated as a Vainshtein mechanism in time domains implies that the extra-degrees of freedom become relevant depending on the direction of time (frame of reference) with respect to the preferred time direction (preferred frame) defined by the St\"uckelberg function $T_0(r,t)$ which contains the information of the extra-degrees of freedom of the theory. In this manuscript, I make the general definition of the Higgs mechanism by analyzing the gauge symmetries of the action and the general form of the vacuum solutions for the graviton field. In general, the symmetry generators depending explicitly on the St\"uckelberg fields are broken at the vacuum level. These broken generators, define the number of Nambu-Goldstone bosons which will be eating up by the dynamical metric in order to become massive.}

\maketitle

\section{Introduction}

If the Einstein-Hilbert action is modified in order to reproduce the accelerated expansion of the universe, in some scenarios, the solar system predictions of General Relativity (GR) are lost due to the additional forces coming from the extra-degrees of freedom of the theory \cite{vDVZ}. This is the case for example of massive gravity, where GR cannot be recovered at solar system scales if the theory under consideration is the linear Fierz-Pauli one \cite{Pauli}. As a consequence of this, some modified gravity theories require the introduction of screening mechanism in order to agree with the solar system observations and other predictions coming from GR \cite{Vainshteinla, Koury}. The most popular screening mechanism is the Vainshtein one, appearing not only inside the non-linear formulation of massive gravity, but also inside most of the scalar-tensor formulations, in particular, inside the Hordenski theory \cite{IntroVainshtein}.    
The Vainshtein mechanism in spatial domains is fundamental for the recovery of the solar system predictions in theories where the Einstein-Hilbert action is modified. The Higgs mechanism at the graviton level has been formulated in \cite{Higgs1} as a Vainshtein mechanism in time domains. This means that the extra-degrees of freedom reproduce the effect of a preferred time direction, analogous to what happens with the spontaneous magnetization in ferromagnetism. Since the extra-degrees of freedom become relevant after the Vainshtein radius ($r_V$) defined in spatial-domains, then it is natural to think of this scale as a phase transition point. However, the formulation of the Higgs mechanism in massive gravity does not require the appearance of a Vainshtein scale in spatial domains. The (Vainshtein) scale will rather appear in time domains in the sense that the relevance of the extra-degrees of freedom will depend on the relative orientation between the ordinary time coordinate defined by observers and the preferred time direction defined by the St\"uckelberg function $T_0(r,t)$. This function contains the information of the extra-degrees of freedom of the theory and it defines a preferred frame of reference. If we examine the diffeomorphism transformations of the action, we can realize that the vacuum is not invariant under the transformations depending explicitly on the St\"uckelberg fields (function). The vacuum solutions, defined as an extremal condition of the massive action if we take the variation with respect to the graviton field, depend explicitly on the total differential of the St\"uckelberg function $dT_0(r,t)$. Then any generator involving explicitly the St\"uckelberg fields, is broken at the vacuum level. Then the vacuum is degenerate and false. In order to get the physical vacuum and then obtain the physical perturbations, it is necessary to shift the perturbations in agreement with the false vacuum. If we rewrite the action in terms of the physical perturbations, the St\"uckelberg functions will appear inside the Christoffel connections, making them equivalent to gauge fields even in the absence of gravity (free-falling frame). In addition, the relative position between the false and physical vacuums, depends explicitly of the St\"uckelberg functions. This means that the Higgs mechanism in massive gravity is in essence non-linear. If the extra-degrees of freedom are absent, then the effect of the St\"uckelberg functions disappears and the false and physical vacuum become the same. In addition, the gauge-portion of the Christoffel connections vanishes. Then the graviton field is in essence massless. An equivalent scenario is given when the direction of time (local frame of reference), selected by the observers is coincident with the direction of the preferred notion of time defined by $T_0(r,t)$. In such a case, the same situation arises and the metric for this particular observers (preferred observers) is massless. This is the notion of Vainshtein mechanism in time domains formulated originally by the author in \cite{Higgs1}.                   

\section{The Vainshtein scale as a function of the St\"uckelberg functions}

In \cite{My paper1}, the Vainshtein scale has been derived as an extremal condition for the massive action at the background level. The Vainshtein scale then marks the distance where the standard gravitational effects are equally important to the effects coming from the extra-degrees of freedom. In general, the scale is time dependent, however, for stationary solutions it is fixed in space. The action in massive gravity theories is defined in general as

\begin{equation}   \label{eq:b1}
S=\frac{1}{2\kappa^2}\int d^4x\sqrt{-g}(R+m_g^2U(g,\phi)).
\end{equation} 
The form of this action is independent of the type of massive gravity theory under consideration. Then issues like ghosts or other pathologies will not be considered relevant for the purposes of this paper. In other words, the methods developed in this manuscript should be considered as general and independent of the theory under consideration. If we ignore the usual Einstein-Hilbert action in (\ref{eq:b1}), then what remains is the potential (massive action) given by

\begin{equation}   \label{eq:b2}
S_{mass}=m_g^2\sqrt{-g}U(g,\phi)=V(g,\phi).
\end{equation} 
Since the graviton mass parameter is a global factor, at the moment of calculating the vacuum conditions, it will be irrelevant. We can define the Vainshtein scale as the algebraic solution of the condition

\begin{equation}   \label{eq:ptm2}
dV(g,\phi)=\left(\frac{\partial V(g,\phi)}{\partial g}\right)_{\phi}dg+\left(\frac{\partial V(g,\phi)}{\partial \phi}\right)_{g}d\phi=0.
\end{equation}
This condition, in unitary gauge is simplified to

\begin{equation}   \label{eq:ptm2sila}
dV(g,\phi)=\left(\frac{\partial V(G)}{\partial G}\right)_{\phi}dG=0,
\end{equation}
where we identify $G$ as the dynamical metric in unitary gauge. Since we introduce all the degrees of freedom inside the objects $G$, then the previous result defines the location of the Vainshtein scale. If the matrix $\left(\frac{\partial V(G)}{\partial G}\right)_{\phi}$ is non-singular, then the simpler condition

\begin{equation}   \label{eq:cond1}
dG_{\mu\nu}=\left(\frac{\partial G_{\mu\nu}}{\partial r}\right)_t dr+\left(\frac{\partial G_{\mu\nu}}{\partial t}\right)_r dt=0,
\end{equation} 
determines the position of the Vainshtein scale which in general is spatial and time dependent. In other words, it can evolve in time.  

\section{Why is the Schwarzschild-de Sitter (S-dS) solution in de Rahm-Gabadadze-Tolley (dRGT) massive gravity with one free-parameter so important?}

In the formulation of the Higgs mechanism at the graviton level in \cite{Higgs1}, the author took the S-dS solution with one free-parameter as the most relevant one. The reason for this, is the fact that this solution has an enhance symmetry which makes the St\"uckelberg function arbitrary. This result was discovered for first time in \cite{Kodama} and extended in \cite{BabiandBrito}, where it was understood that this arbitrariness is in fact an enhanced symmetry for this solution. The enhanced symmetry guarantees that both, the dynamical and the fiducial metric can be transformed independently. Although this fact is not essential in the formulation of the Higgs mechanism at the graviton level, if we switch-off gravity (free-falling frame) in order to simplify the analysis, the arbitrariness of the St\"uckelberg function $T_0(r,t)$ is essential for it to appear in the perturbative vacuum solutions as has been demonstrated in \cite{Higgs1}. In dRGT massive gravity, at the background level, the massive action in this case is given by 

\begin{equation}   \label{eq:utotalpert}
V(g,\phi)=\frac{2+6\alpha(1+\alpha)}{(1+\alpha)^4}.
\end{equation}
Note that the St\"uckelberg function ($T_0(r,t)$) does not appear at the background level for the massive action. Then the arbitrariness of $T_0(r,t)$ does not affect the equations of motion at the background level. In other words, we have a multiplicity of vacuum solutions, all of them defining the same equations of motion for test particles moving around \cite{Komar, HawkingdRGT}. As soon as the perturbations are included, then the functions $T_0(r,t)$ will appear. In such a case, they will affect the behavior of the effective masses defined for the different modes. The masses for the different modes can be found by calculating the second derivatives of the effective potential and evaluating the results at the vacuum level. In the present context, vacuum means the equations of motion of the graviton field (Euler-Lagrange), but ignoring the kinetic terms. If the action is expanded up to quadratic order, then the second derivatives will be independent of the vacuum state. The mass matrix for the different modes is defined as

\begin{equation}   \label{eq:socutethatgirl}
m^{\mu\nu}m^{\alpha\beta}=\frac{\partial^2 V(g,\phi)}{\partial h_{\mu\nu}\partial h_{\alpha\beta}},
\end{equation}
and the vacuum state for the graviton field is defined in agreement with 

\begin{equation}   \label{eq:condilala}
\frac{\partial V(g,\phi)}{\partial h_{\mu\nu}}=0.  
\end{equation}
In general, the components $m^{\mu\nu}$ and the covariant counterparts $m_{\mu\nu}$, will depend on the derivatives of $T_0(r,t)$. Note that the vacuum condition (\ref{eq:condilala}) looks similar to the Vainshtein scale condition defined in eq. (\ref{eq:ptm2}). The difference is that the derivatives in eq. (\ref{eq:condilala}) are taken explicitly with respect to the graviton field. Then it is necessary to expand the action at the perturbative level. On the other hand, the derivatives in eq. (\ref{eq:ptm2sila}) are evaluated with respect to the full dynamical metric and by using the chain rule, the dependence is translated to the coordinates. The perturbations are done around the background metric defined in agreement with

\begin{equation}   \label{eq:metric}
ds^2=-f(Sr)dT_0(r,t)^2+\frac{S^2}{f(Sr)}dr^2+S^2r^2d\Omega^2,  
\end{equation}
with the St\"uckelberg function $T_0(r,t)$ being in principle arbitrary. For the purposes of the present manuscript, the functions $T_0(r,t)$ are considered infinitesimally close to the ordinary time coordinate, such that the portion of the total variations $dT_0(r,t)$, depending on the extra-degrees of freedom are considered infinitesimal.   

\section{The gauge symmetries of the action}

The action (\ref{eq:b1}) in massive gravity is invariant under the following set of transformations 

\begin{equation}   \label{eq:gt}
g_{\mu\nu}\to\frac{\partial f^\alpha}{\partial x^\mu}\frac{\partial f^\beta}{\partial x^\nu}g_{\alpha\beta}(f(x)), \;\;\;\;\;Y^\mu(x)\to f^{-1}(Y(x))^\mu.
\end{equation}
Here the St\"uckelberg function $Y^\mu(x)$, is defined in agreement with 

\begin{equation}   \label{eq:stuefashio2}
Y^0(r,t)=T_0(r,t),\;\;\;\;\;Y^r(r,t)=r.
\end{equation}
In addition, note that the metric $G_{\mu\nu}$ is defined in agreement with 

\begin{equation}   \label{eq:stuefashio}
G_{\mu\nu}=\left(\frac{\partial Y^\alpha}{\partial x^\mu}\right)\left(\frac{\partial Y^\beta}{\partial x^\nu}\right)g_{\alpha\beta},
\end{equation} 
and it is a diffeomorphismn invariant object if we take into account the set of transformations given in eq. (\ref{eq:gt}) \cite{K}. Note that the diffeomorphism invariance of the metric (\ref{eq:stuefashio}) is necessary in order to guarantee the invariance of the massive action, defined in eq. (\ref{eq:b2}), under the same set of transformations. This is the case since in the standard formulation of massive gravity, the massive action depends explicitly on the root square of the partial contractions between the dynamical metric and the fiducial one. In order to be more explicit, we can assume a general dependence for the massive action as follows

\begin{equation}   \label{eq:Vla}
V(g,\phi)=F(g^{\mu\nu}f_{\nu\gamma}),
\end{equation}
with $F$ representing a function depending explicitly of the partial contractions between the dynamical and the fiducial metric. The fiducial metric is in general represented by four scalars $\phi^a$ defined in agreement with

\begin{equation}   \label{eq:Vla2}
f_{\nu\gamma}=\eta_{ab}\partial_\nu\phi^a\partial_\gamma\phi^b.
\end{equation}
The scalars $\phi^a$ are called St\"uckelberg fields. If all the degrees of freedom of the theory are translated to the dynamical metric, then the fiducial one is of Minkowski type and the scalars become trivial coordinates $\phi^a=x^a$ without degrees of freedom. In such a case, the function (\ref{eq:Vla}) becomes

\begin{equation}   \label{eq:Vla3}
V(g,\phi)=F(g^{\mu\nu}f_{\nu\gamma})\to V(G^{\mu\nu}\eta_{\nu\gamma}),
\end{equation}
where the metric $G$ defined in agreement with eq. (\ref{eq:stuefashio}) now appears and $\eta_{\nu\gamma}$ is just the Minkowskian metric. Note that in eq. (\ref{eq:Vla2}), $\eta_{ab}$ is defined in agreement with Minkowski. From the result (\ref{eq:Vla3}), is evident that the diffeomorphism invariance of the massive action $V(g,\phi)$ is only possible if the combination $g^{\mu\nu}f_{\nu\gamma}$ is diffeomorphism invariant, or equivalently, if $G^{\mu\nu}$ is also invariant under gauge transformations. In fact, it is a trivial task to prove that the metric $G$, defined in agreement with eq. (\ref{eq:stuefashio}), is invariant under the transformations (\ref{eq:gt}) \cite{K}. We can define 

\begin{equation}   \label{eq:Vla3lala}
Y^\alpha(r,t)=x^\alpha+A^\alpha,
\end{equation}
where $A^\alpha$ is assumed to be infinitesimally close to zero. In addition

\begin{equation}   \label{alejo1}
A_\mu\to A_\mu+\partial_\mu\phi,
\end{equation}
with $\phi$ defining the scalar component of the St\"uckelberg fields. From the results (\ref{eq:stuefashio2}), it is clear that we have

\begin{equation}   \label{alejo122}
A^0\neq0, \;\;\;\;\;\;\;\;\;\; A^r=0.
\end{equation}
Or equivalently

\begin{equation}   \label{here}
A_r=G_{r0}A^0, \;\;\;\;\;\;\;\;\;\; A_0=G_{00}A^0.
\end{equation}
However, under the re-definitions (\ref{alejo1}), the result (\ref{here}) should be written as

\begin{equation}   \label{here2}
A_r+\partial_r\phi=G_{r0}A^0, \;\;\;\;\;\;\;\;\;\; A_0+\partial_0\phi=G_{00}A^0.
\end{equation}
Note that the metric $G_{\mu\nu}$ also contains the St\"uckelberg fields. Here I will consider these fields perturbatively and then orders higher than two in the St\"uckelberg field expansions are considered negligible. As a consequence of the previous definitions, the diffeomorphism transformations of the theory assuming a flat background metric are defined in agreement with

\begin{eqnarray}   \label{miaumiau}    
\delta_gh_{\mu\nu}=\partial_\mu\zeta_\nu+\partial_\nu\zeta_\mu+\pounds_\zeta h_{\mu\nu},\nonumber\\
\delta_g A_\mu=\partial_\mu\Lambda-\zeta_\mu-A^\alpha\partial_\alpha\zeta_\mu-\frac{1}{2}A^\alpha A^\beta\partial_\alpha\partial_\beta\zeta_\mu-...,\nonumber\\
\delta_g\phi=-\Lambda.
\end{eqnarray}
Note that if we take the background metric to be flat, then the St\"uckelberg functions appearing on it, are equivalent to perturbations around the background and they will be related to the fields appearing in eq. (\ref{miaumiau}). In that sense, eq. (\ref{here}) for example, will include the back-reaction of the St\"uckelberg fields for the definitions of $A_\mu$. The symmetries involved are gauge symmetries and as a consequence, the function $\Lambda$ depends on space-time coordinates. Note that the previous transformations are taken by assuming a flat background metric. The extensions for more general situations is straightforward. What is important for the moment, is to notice that the transformations for the scalar component are of the type $U(1)$. This is important because the scalar component is the one involved in the van Dam-Veltman-Zakharov (vDVZ) discontinuity \cite{vDVZ}.  
      
\section{The symmetries broken at the vacuum level}

The vacuum conditions are given by the result (\ref{eq:condilala}). In general, if the action is expanded up to second order in perturbations, then the vacuum will depend on the total derivative of the St\"uckelberg functions. Then in such a case, we will have results of the form

\begin{equation}   \label{vac}
h_{\mu\nu\;vac}=F_{\mu\nu}(\alpha,\beta,dT_0(r,t)),  
\end{equation}	
where $\alpha$, $\beta$ and $T_0(r,t)$ represent the two free-parameters of the massive action $V(g,\phi)$ and the St\"uckelberg function. Independently of the relation between the parameters $\alpha$ and $\beta$, the explicit dependence of the vacuum results with respect to the St\"uckelberg function implies that the vacuum symmetry is broken with respect to the transformations depending explicitly on $T_0(r,t)$. Given the relations (\ref{eq:stuefashio2}), (\ref{eq:Vla3lala}), (\ref{here}) and (\ref{here2}), this is equivalent to say that the vacuum symmetry is broken with respect to the transformations involving $A_\mu$ and $\phi$. Then even if the action is still invariant under the full set of diffeomorphism transformations defined in eq. (\ref{miaumiau}), the vacuum defined in agreement with eq. (\ref{vac}) is not invariant under the full set of transformations. The function $T_0(r,t)$ represents the preferred time direction of the theory when the extra-degrees of freedom become relevant.     
Then it is clear that the broken generators, are related to the appearance of the extra-degrees of freedom. In this sense, saying that the vacuum symmetries are broken for the gauge transformations involving the fields $A_\mu$ and $\phi$, is equivalent to say that the symmetries under time translations are broken. That this is the case can be perceived from the fact that the set of transformations involving $A_\mu$ and $\phi$ are derived from the general set \cite{K}

\begin{equation}
\delta_gY^\alpha=-\zeta^\alpha(Y),
\end{equation}
which is the infinitesimal version of the transformations given in eq. (\ref{eq:gt}). If we take into account the results given in eq. (\ref{eq:stuefashio2}) for the spherically symmetric solutions derived in \cite{Kodama}, then it is possible to observe that under gauge transformations, the radial component of $Y^\alpha$ transforms trivially as $\delta_gY^r=-\zeta^r(x)=-\delta r$. However, the zero-component has a non-trivial transformation if we compare it with the standard transformation for the ordinary time coordinate given by $\delta_gt=\zeta^t(x)$. Then a symmetry under the ordinary time transformation is not necessarily a symmetry with respect to the time defined in agreement with the St\"uckelberg function and vicerversa. For this reason, in \cite{Higgs1}, the author considered that the broken generators are related to the time coordinate.

\section{The Higgs mechanism at the graviton level}

If we expand the action (\ref{eq:b1}) up to second order, the vacuum will be defined in agreement with eq. (\ref{vac}) after solving the result (\ref{eq:condilala}). The vacuum represented by this solution is false. If we want to obtain the physical vacuum, then it is necessary to shift the perturbations in agreement with the result (\ref{vac}). The method is standard and it was developed in \cite{Higgs1}. Here I show the general result to be

\begin{equation}   \label{shift} 
{h}_{\mu\nu}=\bar{h}_{\mu\nu}+h_{\mu\nu vac},
\end{equation}       
such that the physical vacuum is defined in agreement with the condition

\begin{equation}
\bar{h}_{\mu\nu vac}=0.
\end{equation}
If we rewrite the action (\ref{eq:b1}) in a free-falling frame in terms of the physical perturbations $\bar{h}_{\mu\nu}$, then we obtain the result \cite{Higgs1}

\begin{equation}   \label{lag}
\pounds=\pounds_{EH}+K(dT_0(r,t))+F(v,\alpha)(\Gamma\Gamma)_{gauge}+\bar{V}(g,\phi),
\end{equation}
where the $\Gamma$-matrices are the portion of the Christoffel connections depending explicitly on the St\"uckelberg functions. This portion means that the graviton field has become massive. The term $K(dT_0(r,t))$ corresponds to the portions of the first-order Ricci scalar $R^{(1)}$ that appear after shifting the perturbations in agreement with eq. (\ref{shift}). They represent the Kinetic terms of the gauge fields $\Gamma$. In \cite{Higgs1}, the author did not write explicitly the term $K(dT_0(r,t))$ but it was considered to be included inside the standard Kinetic part $\pounds_{EH}$. Here $v$ is a parameter defined in \cite{Higgs1} and depending on the determinant of the vacuum matrix defined in eq. (\ref{vac}). From the result (\ref{lag}), it is clear that the potential $\bar{V}(g,\phi)$ is modified in agreement with the vacuum shift defined in eq. (\ref{shift}). The shift is such that the new vacuum condition, given by

\begin{equation}    \label{eq:condilala22}
\frac{\partial \bar{V}(g,\phi)}{\partial \bar{h}_{\mu\nu}}=0,
\end{equation}
has as a unique solution the result $\bar{h}_{\mu\nu}=0$. Note that by taking the variations in eqns. (\ref{eq:condilala}) and (\ref{eq:condilala22}), the St\"uckelberg functions $T_0(r,t)$ are kept fixed. However, they will back-react to the final solutions of the perturbations. Keeping fixed the functions $T_0(r,t)$ when we vary the field equations with respect to the field $h_{\mu\nu}$, is a necessary condition in order to satisfy the trivial solution at the vacuum level. Note that the kinetic terms $K(dT_0(r,t))$ in eq. (\ref{lag}), will contain derivatives of order higher than two in a free-falling frame of reference. This interesting scenario will be discussed further in a coming manuscript.   

\section{conclusions}

In this manuscript, I made a formal definition of the Higgs mechanism at the graviton level. The mechanism appears as a Vainshtein mechanism in time domains. The analysis has been based on the gauge-symmetries of the action. The proposed method is general in the sense that it focus on the symmetry aspects of the theory. Then it is valid for any formulation of massive gravity or any other theory including non-derivative terms (in the sense of the metric) interactions in the action in addition to the standard Einstein-Hilbert term. In order to extend to the cases where derivative-term interactions (in the sense of the metric) in the action appear as in the case explored in \cite{MOND, MOND2}, some variations are necessary but the same philosophy would remain. From the previous analysis, it is observed that the symmetries involving explicit dependence on the St\"uckelberg fields are broken at the vacuum level. They are related to the symmetries in the direction of the St\"uckelberg function $T_0(r,t)$. This function represents the preferred time direction of the theory when the extra-degrees of freedom are relevant. When the extra-degrees of freedom are absent, then this functions becomes trivially equivalent to the ordinary time coordinate. Equivalently, an observer defining the time coordinate in agreement with $T_0(r,t)$, will not be able to perceive the effects of the extra-degrees of freedom, because in such a case the gauge transformations with respect to the ordinary time coordinate and those related to the St\"uckelberg function direction will be equivalent. This is the notion of Vainshtein mechanism in time domains. Finally, it is important to notice the appearance of higher derivative terms for the scalar components when the vacuum is re-defined. This important issue, will be explored in a coming manuscript. \\  

{\bf Acknowledgement}
The author would like to thank Shinji Mukohyama, Antonio De Felice and Takahiro Tanaka for the comments around this idea during the "2nd APCTP-TUS workshop on Dark Energy", held at Tokyo University of Science (TUS). The author would like to thank Gianluca Calcagni and Shinji Tsujikawa for useful comments about the methods employed by the author in this paper. The author would like to thank Masahide Yamaguchi and Masaru Siino for the kind invitation to Tokyo University of Science (Titech) and for the subsequent discussions about this idea. Finally, the author would like to thank Ryo Namba and Rene Meyer for the kind invitation to the IPMU at Tokyo University and the subsequent discussions about this result.

\end{document}